\documentclass[%
 aip,
 amsmath,amssymb,
 reprint,%
]{revtex4-1}

\usepackage{graphicx}
\usepackage{dcolumn}
\usepackage{bm}
\usepackage{subfig}

\usepackage[utf8]{inputenc}
\usepackage[T1]{fontenc}
\usepackage{mathptmx}
\newcommand{\etal}{\textit{et al. } }
\usepackage{xcolor}

\newcommand{\iotabar}{\text{$\iota\!\!$-}}
\DeclareMathAlphabet\mathbfcal{OMS}{cmsy}{b}{n}

\begin{document}

\preprint{AIP/123-QED}

\title[]{On the relationship between the multi-region relaxed variational principle and resistive inner layer theory }

\author{A. Kumar}
 \altaffiliation{Arunav.Kumar@anu.edu.au}
\affiliation{ 
Mathematical Sciences Institute, The Australian National University, Canberra, ACT 2601, Australia
}%
 \author{J. Loizu} \altaffiliation{Joaquim.Loizu@epfl.ch}
 \affiliation{  Swiss Plasma Center,
Ecole Polytechnique Federale de Lausanne, CH-1015 Lausanne, Switzerland
}%
\author{M. J. Hole}
\affiliation{ 
Mathematical Sciences Institute, The Australian National University, Canberra, ACT 2601, Australia
}%
\affiliation{%
Australian Nuclear Science and Technology Organisation, Locked Bag 2001, Kirrawee DC NSW 2232, Australia
}%
\author{Z.S. Qu}
\affiliation{ 
Mathematical Sciences Institute, The Australian National University, Canberra, ACT 2601, Australia
}%

\author{S. R. Hudson }
\affiliation{%
Princeton Plasma Physics Laboratory, PO-415, New Jersey 08540, United States of America
}%
\author{R. L Dewar}
\affiliation{ 
Mathematical Sciences Institute, The Australian National University, Canberra, ACT 2601, Australia
}%

\date{\today}

\begin{abstract}
  We show that the variational energy principle of multi-region relaxed magnetohydrodynamic (MRxMHD) model can be used to predict finite-pressure linear tearing instabilities.  In this model, the plasma volume is sliced into sub-volumes separated by ``ideal interfaces'', and in each volume the magnetic field relaxes to a Taylor state where the pressure gradient $\nabla p =0$. The MRxMHD model is implemented in the SPEC code so that the equilibrium solution in each region is computed while the preserving force balance across the interfaces. As SPEC computes the Hessian matrix (a discretized stability matrix), the stability of an MRxMHD equilibrium can also be computed with SPEC. In this article, using SPEC, we investigate the effect of local pressure gradients and the $\nabla p =0$ in the vicinity of the resonant surface of a tearing mode. For low beta plasma, we have been able to illustrate a relationship between the resistive singular layer theory [Coppi \etal(1966) Nucl. Fusion 6 101, Glasser \etal The Physics of Fluids 18, 875-888 (1975)], and the MRxMHD model. Within the singular layer, the volume-averaged magnetic helicity and the flux-averaged toroidal flux are shown to be the invariants for the linear tearing modes in SPEC simulations. Our technique to compute MRxMHD stability is first tested numerically in cylindrical tokamak and its application in toroidal geometry is demonstrated. We demonstrate an agreement between the stability boundary obtained with SPEC simulation and the resistive inner layer theories. 
\end{abstract}

\maketitle
\section{Introduction}

\subsection{Theoretical review}
One of the most fascinating aspects of the MHD events in these plasmas is the existence of plasma oscillations that results in sawteeth due to current-induced instabilities that are either internally or externally caused by electron cyclotron resonance heating \citep{Chapman2010ControllingSO}. Such explosive events can lead to magnetic reconnection as the magnetic energy is converted locally into particle heat and kinetic energy via, the mechanism of effective magnetic dissipation that allows for a change of magnetic field line connectivity. More generally, magnetic reconnection is also the heart of well-know studies for, solar flares \cite{Shibata2011}, geomagnetic sub-storms \cite{10.3389/fspas.2020.604750} and electron dynamics in tokamaks \cite{PhysRevLett.115.215001}. 
\par
Resistive tearing MHD instabilities are of considerable fundamental as well as practical importance in the study of sawteeth oscillations. Coppi-Greene-Johnson (CGJ)\citep{Coppi_1966} and Glasser- Greene-Johnson (GGJ)\citep{GGJ1976} demonstrated that the finite pressure has a stabilizing effect on tearing instabilities. In their MHD theory, the resistive modes were analyzed
using the boundary layer theory when the boundary layers occur near the rational surfaces. In other words, the plasma can be analyzed in two regions: an
“outer region” where the plasma is ideal and an “inner layer” where non-ideal dissipation is significant. The resistive inner-layer equations in cylindrical geometry were derived by CGJ, and the extension of these equations in general toroidal geometry were derived by GGJ. In a pressureless plasma \citep{NEWCOMB1960232}, the condition
for instability can be understood as $\Delta' > 0$. In a plasma with pressure, an instability can be possible only if $\Delta' > \Delta_{crit}$, where $\Delta'$ is the stability parameter measuring the free energy available for the mode, and $\Delta_{crit}$ is a positive threshold value. Implications of this stabilizing effect of finite pressure have been explored in several numerical studies of the resistive stability of tokamaks \citep{glasser2016,glasserandwang}. As a result, this approach is extensively applied to study tearing instabilities, and has attractive features such as high numerical efficiency and clear separation of the physics between the ideal outer and the resistive inner regions. Consequently, the dependence of tearing mode stability on the pressure-gradient $\nabla p$ at the rational surface is well understood. 

In the absence of $\nabla p$ at the resonant surface,
two approaches to solving this problem have been reviewed in Ham \etal \citep{Ham_2012}. Firstly one can deduce $\Delta'$ from the tearing
mode growth rate $\omega$, calculated by a resistive MHD code using the known dispersion relation for the resistive MHD model. Alternatively, one can use a resistive MHD code to obtain a set of basis functions from which $\Delta'$ can be constructed. This
involves a set of fully reconnected solutions (\textit{i.e.} continuous at
the relevant resonant surfaces and thus containing the large
solution in the sense of Newcomb \citep{NEWCOMB1960232} ) and a set of small
solutions, again in the sense of Newcomb, emerging from
the various resonant surfaces which can then be combined to
satisfy the appropriate boundary conditions at the magnetic
axis and plasma edge and used to deduce $\Delta'$.
However both approaches have difficulties when there is
a non-zero pressure gradient and favourable average curvature at a resonant surface. This is known as the ‘Glasser effect’ \citep{GGJ1976}. Naturally, the pressure flattening may arise in experiments, say,  due to the presence of magnetic islands at a resonant surface when rapid transport along the field lines short circuits the cross-field confinement. Ham \etal\citep{Ham_2012} described an artificial pressure flattening function at the rational surfaces (which tends to remove the Glasser effect), and its relationship has been established with the calculation of $\Delta'$. Alongside, Bishop \etal \citep{Bishop_1991} discussed a localized pressure flattening perturbation at the resonant surface in order to assess the sensitivity of $\Delta'$ to such effects. Indeed, the modelling of this in linear tearing theory by including anisotropic thermal transport in the governing equations can significantly modify the Glasser
dispersion relation \citep{lut_2001,L_tjens_2001}, yielding the form of the natural diffusion length-scale as,
\begin{eqnarray}
    {w}_{d} =2\sqrt{2}\left(\frac{\chi_{\perp}}{\chi_{||}}\right)^{1/4} \left(\frac{r_{s}R_{0}}{n\,s}\right)^{1/2}.
\end{eqnarray} Here, $R_{0}$ is the major radius, $r_{s}$ is the radial location of a rational surface, $n$ is the toroidal mode number, $s = r q'/q$ is the positive
equilibrium magnetic shear and $q$ is the safety factor and $\chi_{\perp}$ and $\chi_{||}$ are the perpendicular and parallel transport thermal diffusivities, respectively. Ham \etal \citep{Ham_2012} showed that $ \delta_{L}< w_{d} $ where $\delta_{L}$ is the characteristic length-scale within the resistive  singular layer . For the anisotropic plasmas\citep{richard} ($\chi_{\perp}< \chi_{||}$) , one could see that $ \delta_{L}<w_{d}< 2\sqrt{2}(r_{s}R_{0}/n \,s)^{1/2} $.  Thus, the upper bound of $\delta_{L}$ has no dependence on plasma resistivity. 


\par
\subsection{Multi-Region relaxed MHD model}
In this article, we focus on determining pressure-induced tearing instabilities using the variational energy principle of multi-region relaxed MHD (MRxMHD). The MRxMHD variational principle is a generalization of the global Taylor's relaxation conjecture to the partial relaxation, to understand physical mechanism of magnetic reconnections. Before the Taylor's relaxation conjecture, the process of magnetic relaxation in a low-resistivity plasma was envisaged by Woltjer \citep{woltjer_1958,woltjer_1958b}. Taylor's conjecture postulates that, in a turbulent low-density and resistivity plasma, the global magnetic helicity is well conserved. This evidently leads to Beltrami fields, of the form $\nabla \times {\bf B}= \mu {\bf B}$ with $\mu =const.$ and $\nabla p =0$, which provided an explanation to the field structures observed in Reverse-field pinches (RFPs). In the late 2000s, Dewar, Hole \etal \citep{hole_hudson_dewar_2006} generalised the Taylor's theory and proposed the MRxMHD variational energy principle, which is also referred as stepped-pressure MHD model. 

\par 
In MRxMHD, the whole plasma $\Omega$ is partitioned into the discrete number of Beltrami relaxed plasma regions, $\Omega_{l}$'s, such that the pressure, $p_{l}=const.$ and $\nabla p_{l}=0$ within these regions. Each relaxed region is then bounded by freely variable toroidal interfaces, on the outer edge by the interface $I_l$ and on the inner edge by $I_{l-1}$ that are assumed invariant ideal surfaces during the minimization of MRxMHD energy. The multi-region relaxed-MHD energy principle \cite{hole_hudson_dewar_2006},
minimizes the total potential and thermal energy under an invariant topological constraints, known as magnetic helicity $K_l$, which takes a form \citep{hudson_2012} \begin{eqnarray}\label{eq.6}
\mu_0F = \sum_{l=1}^{N_v} \left[\int_{\Omega_l} \left(\frac{p}{\gamma-1}+\frac{B^2}{2}\right) d^{3}\tau - \frac{\mu_l}{2} ( K_l-K_{l,0}) \right].
\end{eqnarray} 
Here, the magnetic helicity, identifies as the volume-preserving invariant quantity under the gauge transformation of vector potential, ${\bf A}\rightarrow {\bf A}+\nabla \Xi$ where $\Xi$ is a single-valued gauge potential, alongside the toroidal and poloidal magnetic fluxes. In each $\Omega_l$, the mass and entropy constraints yields an isentropic, ideal-gas constraint, $p_{l} V_{l}^{\gamma}= c_{l}$ where $V_l$ is the volume of $\Omega_l$ and $c_{l}$ is a constant. The volume $\Omega_l$ enclosed by ``\textit{\textit{ideal interfaces }}'', are constrained to have helicity $K_{l,0}$, the poloidal flux $\Delta\psi_{p,l}$, and the toroidal flux $\Delta\psi_{t,l}$. This theory unifies the ideal MHD energy principle and Taylor's relaxation conjecture \cite{PhysRevLett.33.1139} by allowing less-restrictive class of variations in comparison to ideal MHD. These variations allow magnetic reconnection to form islands and chaotic fields. MRxMHD shows no explicit dependence on non-ideal dissipation parameters. 
 \par

    To numerically access the extremizing states of MRxMHD plasmas, the Stepped-Pressure Equilibrium Code (SPEC) \cite{hudson_2012,Hudson_2020} was developed.  SPEC uses  pseudo Galerkin method with a Fourier-Galerkin discretization, and operates in slab, cylindrical, and toroidal geometry. For fixed-boundary simulations, SPEC requires as inputs the plasma boundary and the $N_{v}$ number of Taylor relaxed volumes, the enclosed poloidal $\Delta \psi_{p,l}$ and toroidal flux $\Delta \psi_{t,l}$, and magnetic helicity $K_{l,0}$ in each volume $\Omega_l$ \textit{i.e.} $\{p,\Delta \psi_{p,l},\Delta \psi_{t,l},K_{l,0}\}$. Alternatively, if the helicity multiplier $\mu_l$ or parallel current is given, then the equilibrium can be described by $\{p,\Delta \psi_{p,l},\Delta \psi_{p,l},\mu_l\}$. Then, as a part of the energy minimization process the geometries of the ideal interfaces are varied to ensure that force balance is achieved across each barrier. 
    \par The linear stability of MRxMHD equilibria in SPEC can also be assessed
     to analyze the magneto-hydrostatic instabilities. It is shown that the linear stability analysis of
MRxMHD can reproduce both ideal and resistive MHD stability results \citep{Hole_2009,loizu2019,Kumar_2021,Kumar_2022}. General derivations of the SPEC-stability
matrix in toroidal geometry are discussed in Hennenberg \etal \citep{henneberg_hudson_pfefferl_helander_2021} and Kumar \etal\citep{Kumar_2022}. 
\par 
For MRxMHD, a particular effort has been made to clarify the relationship with the outer resistive boundary layer stability condition, that is, $\Delta'$ for the case of pressure-less slab and cylindrical tokamak plasma. Using SPEC, in slab geometry, Loizu and Hudson \citep{doi:10.1063/1.5091765} found that the variational principle of MRxMHD and the corresponding stability boundary is in exact agreement with linear tearing
mode condition $\Delta'$, for $\tilde{\delta}/L<0.2$ where $\tilde{\delta}$ is the arbitrarily small thickness (or width) of the resistive current sheet layer and $L$ is the length of the current sheet along $y$- direction. For finite pressure cylindrical plasma, CGJ showed that a pressure gradient within the resistive layer can drive a tearing-type instability, and GGJ later showed that in toroidal geometry, where the average curvature is
favourable, tearing modes can be strongly stabilized by pressure gradient effects within the resistive layer. To our knowledge, there haven't been any studies conducted on the pressure-induced tearing modes for MRxMHD plasmas. The question addressed in this article is, what happens to the
MRxMHD resistive layer as a result of tearing instability with finite-pressure ? We investigate these mode stability with our compressible MRxMHD stability model, and clarify its applicability regimes. This goal is achieved by as follows: investigating a mode stability in cylindrical and tokamak geometry, where the role of the resistive volume layer in MRxMHD can be easily quantified. \par An another motivation behind utilizing the MRxMHD stability
using SPEC is stellarator optimization. In recent years, SPEC has emerged as one of the MHD tool for stellarator design and optimization studies \citep{matt2021,elizabethmatt,baillod2022}. If the stability to tearing modes and the ideal MHD stability of interfaces can be computed for free during equilibrium parameter scans, then the computational cost of stellarator optimization could be reduced. 
\par

This article is structured as follows. Section \ref{sec2a} outlines the stability condition between the CGJ resistive layer theory and MRxMHD model, in cylindrical geometry.  We extend our generalized expression for the Hessian matrix of Kumar \etal \citep{Kumar_2021} to account for finite compressibility $\gamma=5/3$. Sec. \ref{sec2b} investigates the role of resistive inner layer and compares the stability boundary obtained with SPEC Hessian and the CGJ model. Sec. \ref{sec3a} extends the work to toroidal geometry and describe the stability condition between the GGJ resistive layer theory and MRxMHD model. Sec. \ref{sec3b} provides a comparison study of marginal stability prediction of SPEC stability with the GGJ model, and highlight the significance of MRxMHD energy principle to predict modified tearing mode. Finally, Sec.\ref{sec4} discusses the conclusion and identifies future work.

\section{Resistive interchange mode in cylindrical tokamak}\label{sec2}

In this section, we examine the variational energy principle of MRxMHD for resistive interchange modes in a cylindrical tokamak, and compare it with the Coppi, Greene and Johnson compressible resistive layer model. Much of the physical picture underlying this instability has been well known \citep{prl,jayakumar}. The major result of this section lies with the establishment of a clear relationship between the resistive inner layer of CGJ and MRxMHD model, in the cylindrical geometry. 

\subsection{Stability condition between CGJ and MRxMHD model}\label{sec2a}
    In the CGJ model, all the dynamics of the tearing mode is contained in the linearizied set of resistive MHD equations, which can be written as, 
\begin{eqnarray}\label{eqr.1}
   \rho\omega^{2}\boldsymbol{\xi} = (\nabla \,\times\, {\bf b}) \times\, {\bf B} + {\bf J} \, \times\, {\bf b} + \nabla( \gamma p \nabla\cdot\boldsymbol{\xi} + \boldsymbol{\xi} \cdot \nabla p),
\end{eqnarray}
 \begin{eqnarray}\label{eqr.2}
    {\bf b} - \eta\nabla^{2} {\bf b} = \nabla \, \times \,(\boldsymbol{\xi} \,\times\, {\bf B}). 
 \end{eqnarray} where $\boldsymbol{\xi}$ and $\bf b$ denotes the perturbed velocity and magnetic field, respectively and $\eta$ is the plasma resistivity. Fruth \etal \citep{frs1963} showed that the approximate balance between the curvature force driving the interchange mode and restoring magnetic forces within the resistive singular layer requires that $qr^{2} \sim \eta$.
 
 \par Consider the coordinates ($r,\theta,\phi$) such that the
equilibria depends only on the radius $r$. We non-dimensionlize all quantities: scaling length to the plasma-wall boundary (such that $a=1$ )  and  the magnetic field to its axis $r=0$ such that $B_{z}(0) = 1$. In cylindrical tokamak ordering \citep{iacanno_1994}, the CGJ described the resistive inner layer equations as,
 \begin{eqnarray}
     b_{r}^{''}&=& Q\,({b}_{r} - r \,{\xi}_{r}),\label{eqn6}\\
    Q^2 {\xi}_{r}^{''} &=& Q\,r^{2}{\xi}_{r} - D_{s}{ Y} - Q_{r} \,{b}_{r},\\
{Y}^{''}&=& \left(Q+\frac{Q}{\beta} +\frac{r^2}{Q}\right){Y} -  \left(Q+\frac{Q}{\beta} - \frac{Q\,S}{D_s}\right){\xi}_{r} - \frac{r}{Q}{b_{r}},\label{eqn8}    
 \end{eqnarray} where the ${b_{r}}$, $ {\xi}_{r}$ and $ Y$ denote the radial component of, the perturbed magnetic field, the electrostatic potential/displacement vector and the perturbed pressure along the equilibrium magnetic field, respectively. Here, $Q = \omega \delta_{L}^{2}/\eta$ and $\delta_{L}=\eta^{1/3}\rho^{1/6}\left(\frac{qr_{s}}{mq' B_{\theta}}\right)^{1/3}$, defined as the characteristic resistive  thickness of the inner layer, where $\rho$ is the mass density, $m$ the poloidal mode number, and $B_{\theta}$ the azimuthal component of the equilibrium field.  The other components such as magnetic shear $S$, Suydam's paramter $D_{s},$ and $\beta$ depends upon plasma equilibrium quantities. The solutions within the resistive layer then match with the inertia-free outer layer stability condition $\Delta'$ (Fruth \etal \citep{frs1963}), using the asymptotic matching technique. For finite-compressibility, the dispersion relation for the system of Eqns.\eqref{eqn6}-\eqref{eqn8} is obtained from the matching condition, given as, $a\,\Delta'=\, \Delta_{}(Q)$ with $Q$ being complex. For the particular choice of parameters, an equilibrium is resistive interchange unstable only if $\Delta'$ exceeds a critical value, $\Delta_{crit} > 0$ (see Eqn.(9) of Ham \etal \citep{Ham_2012}), in the vicinity of the resonant location. That is, the stability occurs when $\Delta' < \Delta_{crit}$.
 
 \par
To examine the stability threshold of resistive interchange mode in the vicinity of resonant singular layer, we introduce a localized stability parameter $Z(\delta_{L}^{CGJ}) $ defined as \begin{eqnarray}\label{eq.8}
Z(\delta_{L}^{CGJ})=\frac{a\Delta'}{\Delta_{crit}}.
\end{eqnarray} An instability will occur if $Z(\delta_{L}^{CGJ}) > 1$ for $\Delta_{crit} > 0$,  such that the marginal stability threshold is determined when $Z(\delta_{L}^{CGJ}) = 1$.  By doing this, we will be able to bring out the relationship between the resistive layer of CGJ and the MRxMHD model. 
 \par
\begin{figure}[ht]
\centering
{\includegraphics[width=9.0cm]{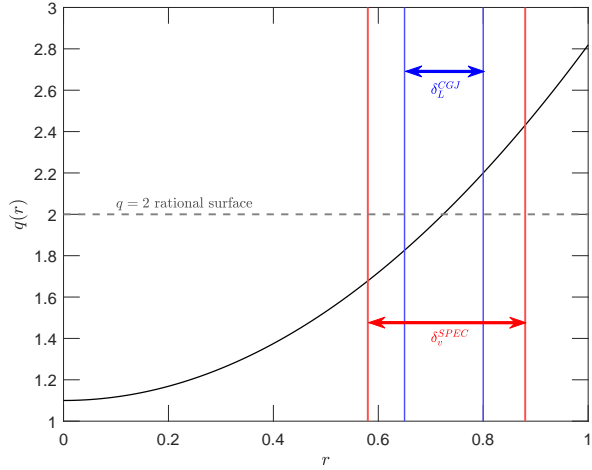}}
\hspace{0mm}
\caption{ A schematic sketch of  $\delta_{v}^{SPEC}(r)$ and $\delta_{L}^{CGJ}(r)$ in the vicinity of $q=2$ rational surface (dashed grey line). The $\delta_{v}^{SPEC}$ denotes the characteristic radial width of a resonant volume in the vicinity of singular surface (where $\nabla p =0$ and $p=const.$).  The $\delta_{L}^{CGJ}$ denotes the characteristic resistive thickness of the singular layer where $\nabla p \neq0$. Note that, $\delta_{v}^{SPEC}$ and $\delta_{L}^{CGJ}$ are two independent quantities. Here, the minor radius $a=1$ and the safety factor profile is defined as $q(r)=1.1(1+(r/0.8)^{2})$. Figure not to scale.  } 
\label{delt_cyc}

\end{figure}

In this article, we restrict our stability consideration only to the vicinity of singular surface where the \textit{rotational transform} $\iotabar=\iota/2\pi=n/m=1/2$ is rational. Traditionally, the linear and non-linear tearing mode layer theories predict a stabilizing effect arising from local pressure gradients at the resonant surface coupled to favourable average curvature \citep{hegna,peng}. In case of MRxMHD theory, the pressure gradient $\nabla p$ is considered to be zero in the vicinity of resonant rational surface, to circumvent the Pfirsch–Schlüter current, which takes form of a $1/x$ singularity. Thus in order to satisfy this condition, for a given characteristic radial width of a volume $\delta_{v}^{SPEC}$ ($<\!\!<$ the plasma minor radius, $a$) the resonant rational surface where ${\bf k}\cdot {\bf B}=0$ must falls within the Taylor relaxed volume. Thus, the $\delta_{v}^{SPEC}$ is understood as an user-defined parameter in SPEC such that $0<\delta_{v}^{SPEC}<\!\!\!<a$. In Figure \ref{delt_cyc}, we have shown a schematic sketch of $\delta_{v}^{SPEC}$ and $\delta_{L}^{CGJ}$ as a function of $r$ in the vicinity of the $q=m/n=2/1$ rational surface (dashed grey line). 
\par 
The majority of the pressure gradients are localized on the ``\textit{ideal interfaces }''. As a consequence, the ``\textit{ideal interfaces }'' enclosed adjacent to a resonant volume must have irrational \textit{rotational transform} \iotabar. This condition is also extremely crucial.  If an ``\textit{ideal interface}'' persists of a \textit{rotational transform} $\iotabar \in \,\mathbf{Q}^{+}$ and the pressure jump is non-zero, then that surface can be unstable to localized ideal modes driven by surface currents \citep{surface}. 
\par
\par
Finite pressure jumps are allowed at the irrational surfaces, and can be interpreted as non-resonant, Kolmogorov, Arnold and Moser (KAM) surfaces \citep{hudson_kraus_2017}. 
To obtain an irrational \textit{rotational transform} on an ``\textit{ideal interface}'' or a surface (which satisfies the condition ${\bf B}\cdot{\bf n}=0$), Greene and Mackay \citep{Gre79,MacKay_1992,MACKAY198392} provided  very insightful, precise methods to determine the existence of a given irrational surface (and it only really makes sense to describe an invariant surface by the \textit{rotational transform} \iotabar). The existence of a given irrational surface is also closely related to the stability of nearby periodic orbits. That is, when $\iotabar$ is irrational,
a single field line ergodically covers the flux surface, and
the surface is referred as an irrational surface. 
\par
We will now explain how to establish an MRxMHD equilibrium and its stability with SPEC. To find an MRxMHD equilibrium, in each $\Omega_{l}$, SPEC requires the pressure $p_{l}$, the enclosed poloidal $\Delta \psi_{p,l}$ and toroidal flux $\Delta \psi_{t,l}$, and the magnetic helicity $K_{l}$ as input parameters \textit{i.e.} $\{ p,\Delta \psi_{t},\Delta \psi_{p},K\}_{l=1,2...N_{v}}$. So, at first, for the special case of the relaxed volume containing the resonant rational surface, we let the resonant volume to have an arbitrary radial width, denoted by $\delta_{v}^{SPEC}$. The input parameters $K_{l}$, $\Delta \psi_{t,l}$, $\Delta \psi_{p,l}$ within each $\Omega_{l}$, are determined by discretizing the volume-averaged magnetic helicity $\langle K_{}\rangle$ and the flux-averaged toroidal and poloidal fluxes ($\tilde{\psi}_{tw}$ and $\tilde{\psi}_{pw}$) profiles over the required number of volumes $N_{v}$. These averaged quantities are obtained and evaluated from the equilibrium profiles which are considered for the simulation. Thus, the size of $\delta_{v}^{SPEC}$ is parameterized by the $K_{l}$ and the enclosed fluxes. Note that the $<K_{}>$ is normalised to the total helicity and $\tilde{\psi}_{tw}$ varies between 0 and 1. 
Then, adjacent to the resonant volume where we determine the KAM surfaces, we represent them by the functional form of a co-ordinate surface. To adapt such surfaces in SPEC cylindrical geometry, their co-ordinate geometry can be constructed as, $ {\bf x}_{l}(\theta, \zeta) = R(\theta,\zeta) \cos \theta  \hat{\bf i}+ R(\theta,\zeta) \sin \theta  \hat{\bf j} + \zeta \hat{\bf k}$, where the toroidal angle, $\zeta$, is identical to the cylindrical angle, $\zeta \sim \phi$. The $R(\theta,\zeta)$ is understood to be in the Fourier summation 
of $R(\theta, \zeta) =\sum_{i} R_{i} \cos(m_{i}\theta - n_{i}N_{p}\zeta)$ where $m_i$ and $n_i$ are $i^{th}$ poloidal and toroidal harmonics and $N_p$ is the field periodicity.

In cylindrical geometry SPEC, the stability of any MRxMHD equilibrium can be assessed by considering the infinitesimal variation of the interface force balance term, $f_{l}=-[[p+B^{2}/2\mu_{0}]]_{l}{\bf n}_{l}$, with respect to the perturbation of interface geometry ${\bf x}_{l}$ \cite{Kumar_2021}. Here, ${\bf n}_{l}$ denotes the unit normal to the interface ${\bf x}_{l}$. This form of change is numerically interpreted as the Hessian matrix, which can be written as 
\begin{eqnarray}\label{eq.hess.45}
    {\mathbf H}_{j,k,l,{l'}}&=&\frac{\delta}{\delta {\bf x}_{l',k}} \left(\delta F/\delta {\bf x}_{l,j}\right),
\end{eqnarray} where $j$ and $k$ are defined as dummy variable for the Fourier harmonics for reader's clarity, with $N_{m,n}$ being the total number of Fourier modes, and $l$ and $l'$ represent the different interface labels. The above equation is expanded as the Fourier summation $\partial {F}_l/\partial {\bf x}_{l_{}}=\sum_i\partial {F}_l/\partial {\bf x}_{l_{},i}\cos(m_i\theta-n_iN_{p}\zeta)$ and $ {\bf x}_l=\sum_i {\bf x}_{l,i}\cos(m_i\theta-n_iN_{p}\zeta)$. When the matrix $\mathbf{H} $ is evaluated at fixed magnetic helicity and enclosed fluxes (toroidal and poloidal), its eigenvalues provide information about the stability corresponds to each Fourier mode harmonics $m_{i},n_{i}$'s. To take account of finite compressibility ($\gamma=\frac{5}{3}$) in SPEC, the pressure variations can be computed adiabatically, as the change in corresponding volume $V_{l}$ of relaxed plasma volumes $\Omega_{l}$, that is \begin{eqnarray} \label{pre_Var}
    \frac{\delta p_{l}}{p_{l}} = -\gamma\frac{\delta V_{l}}{V_{l}},
\end{eqnarray}  where the $\delta V_{l}= (\partial V_{l}/\partial R_{l})\delta R_{l}$.
 An expression to compute the volume $V_l$ which is enclosed by the $l^{th}$ and $({l-1})^{th}$ interface can be obtained by an integral form of
\begin{eqnarray}
 V_{l} &=& \int_{{ \Omega_{l}}} d^{3}\tau = \frac{1}{2}\int_{{ \Omega_{l}}} \; \nabla \cdot {\bf x}_{l}\,\, d^{3}\tau = \frac{1}{2} \int_{{ \partial \Omega_l}} \; {\bf x}_{l} \cdot d{\bf S}, \\
 &=& \frac{1}{2} \int_{0}^{2\pi} d{\theta} \int_{0}^{2\pi/N_{p}} d{\zeta} ({\bf x}_{l} \cdot {{\bf e}_{\theta}} \times 
{{\bf e}_{\zeta}}),\\
   &=&  \frac{1}{2} \int_{0}^{2\pi}\!\!\!d{\theta} \int_{0}^{2\pi/N_{p}} d{\zeta} \; R_{l}^{2},
 \end{eqnarray} where we have used $\nabla \cdot {\bf x}_l = 2$ (because it is 2D), and have assumed that the domain is periodic in the angles. The above equation is understood as a summation of the Fourier harmonics as 
\begin{eqnarray}
\begin{aligned}
    V_{l}  & =  \frac{1}{2} \; \frac{4\pi^{2}}{N_{p}} \; \sum_{i}\sum_{j}\,R_{l,i}R_{l,j}\,\oint\!\!\!\!\oint \!\! d\theta d\zeta \;\cos\alpha_{i}\,\cos\alpha_{j} ,
\end{aligned}
\end{eqnarray}
where $i^{th}$ and $j^{th}$  are the Fourier harmonics of $R_{l}$, $\alpha_{i} = m_{i}\theta - n_{i}\zeta$ and $\alpha_{j} = m_{j}\theta - n_{j}\zeta$. The required partial derivative $\frac{\partial V_{l}}{\partial R_{l,i}}$, with their trigonometrical quantities can be obtained as 
 \begin{align}
  \frac{\partial V_{l}}{\partial R_{l,i}} = \frac{1}{2} \; \frac{4\pi^{2}}{N_{p}} \; \sum_{i}\sum_{j}\,R_{l,j}\,[2\cos(\alpha_{i}-\alpha_{j})+\,2\cos(\alpha_{i}+\alpha_{j})].
    \end{align} 

The symmetry of $\bf H$, means that all its eigenvalues are real numbers \cite{sur1995}. Using the principle axis theorem \cite{strang09}, the quadratic form $\delta {\bf{x}}^T \cdot {\bf H} \cdot \delta {\bf{x}}$ can be condensed as
 \begin{eqnarray}
    \delta {\bf{x}}^T \cdot {\bf H} \cdot \delta {\bf{x}}= \sum_{j}\lambda_{j}v_{j}^{2},
 \end{eqnarray} where the $\lambda_{j}$ is the eigenvalue of $\bf H$ and $v_{j}$ is the corresponding eigenvector for $j=(1,2,3.....,N_{mn}$). The stability of an equilibrium can be predicted from the sign of eigenvalue $\lambda_{j}$ that is, if there exist a $j$ such that $\lambda_j < 0$ then an equilibrium is said to be unstable and if all the $\lambda_j > 0$                                                                                                                                                                                                                                                                                                                                                                                                                                                                                                                                                                                                                                                                                                                                                                                                                                                                                                                                                                                                                                                                                                                                                                                                                                                                                                                                                                                                                                                                                                                                                                                                                                                                                                                                                                                                                                                                                                                                                                                                                                                                                                                                                                                                                                                                                                                                                                                                                                                                                                                                                                                                                                                                                                                                                                                                                                                                                                                                                                                                                                                                                                                                                                                                                                                                                                                                                                                                                                                                                                                                                                                                                                                                                                                                                                                                                                                                                                                                                                                                                                                                                                                                                                                                                                                                                                                                                                                                                                                                                                                                                                                                                                                                                                                                                                                                                                                                                                                                                                                                                                                                                                                                                                                                                                                                                                                                                                                                                                                                                                                                                                                                                                                                                                                                                                                                                                                                                                                                                                                                                                                                                                                                                                                                                                                                                                                                                                                                                                                                                                                                                                                                                                                            `` then an equilibrium is said to be stable.  These eigenvalues hereafter regarded as $\lambda^{m,n}_{SPEC}$, from $\bf{H}$ are evaluated numerically using the SPEC-Hessian calculation. 
 \par
 
 {To compute the stability condition of CGJ with SPEC}, we conform to their notation and express the smallest negative eigenvalue (normalized to its maximum value) referred as $min\{\lambda^{m,n}_{SPEC}\}$ in terms of the stability parameter $Z(\delta_{v}^{SPEC})$, defined by ,\begin{eqnarray}\label{deltaSPEC}
    Z(\delta_{v}^{SPEC})= 1- min\{\lambda^{m,n}_{SPEC}(\delta_{v}^{SPEC})\}.
\end{eqnarray}
 Therefore, the stability conditions of $Z(\delta_{v}^{SPEC})$ can be interpreted as, if there exists  $\lambda^{m,n}_{SPEC}(\delta_{v}^{SPEC}) < 0$, then an instability occurs when $Z(\delta_{v}^{SPEC}) > 1$, and the stability is determined for $Z(\delta_{v}^{SPEC}) < 1$.
\subsection{Equilibrium and instability threshold }\label{sec2b}
To clarify the concepts described in previous section, we discuss the resistive interchange instability of a cylindrical equilibrium considered in Izzo \etal \citep{izzo1985}. The equilibrium of interest is described by the pressure profile \begin{eqnarray}
    p(r) = p_{0}(0.001+0.028\,r^2 -0.059\, r^4 + 0.03\, r^6),
\end{eqnarray} and the safety factor profile as
\begin{eqnarray}\label{eqn.q.cycl}
    q(r) = q_{0}(1+(r/0.8)^2),
    \end{eqnarray} where $q_{0}=1.6$, the aspect ratio $A=5$ and $ 0 \leq p_{0} \leq1$. For the mode perturbation, $m =2$, $n= 1$ this equilibrium is always tearing unstable. The physical motivation for using the equilibrium with the pressure gradient reversed at the resonant layer is to simulate the effects of
good average curvature. Izzo \etal \citep{izzo1985} investigated what happens to the resistive tearing mode at a fixed resistivity as the pressure parameter $p_{0}$ is gradually increased. 
\par 

\begin{figure}[ht]
\centering
{\includegraphics[width=9.0cm]{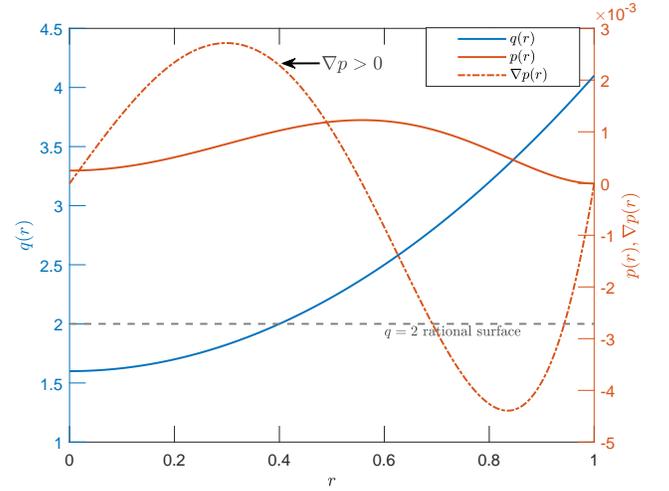}}
\hspace{0mm}
\caption{ Equilibrium configuration : Left axis - the safety factor profile $q(r)$ (in blue curve) as a function of $r$. The $q=2$ rational surface denoted in light gray line, Right axis - the pressure profile $p(r)$ (in red curve) for $p_{0}=0.25$ and the pressure gradient $\nabla p$ profile (in dashed red curve), as a function of $r$. 
  } 
\label{q_cycl}

\end{figure}
 
\begin{figure}[ht]
\centering
\subfloat[\label{cyl_tearing}]{\includegraphics[width=9.0cm]{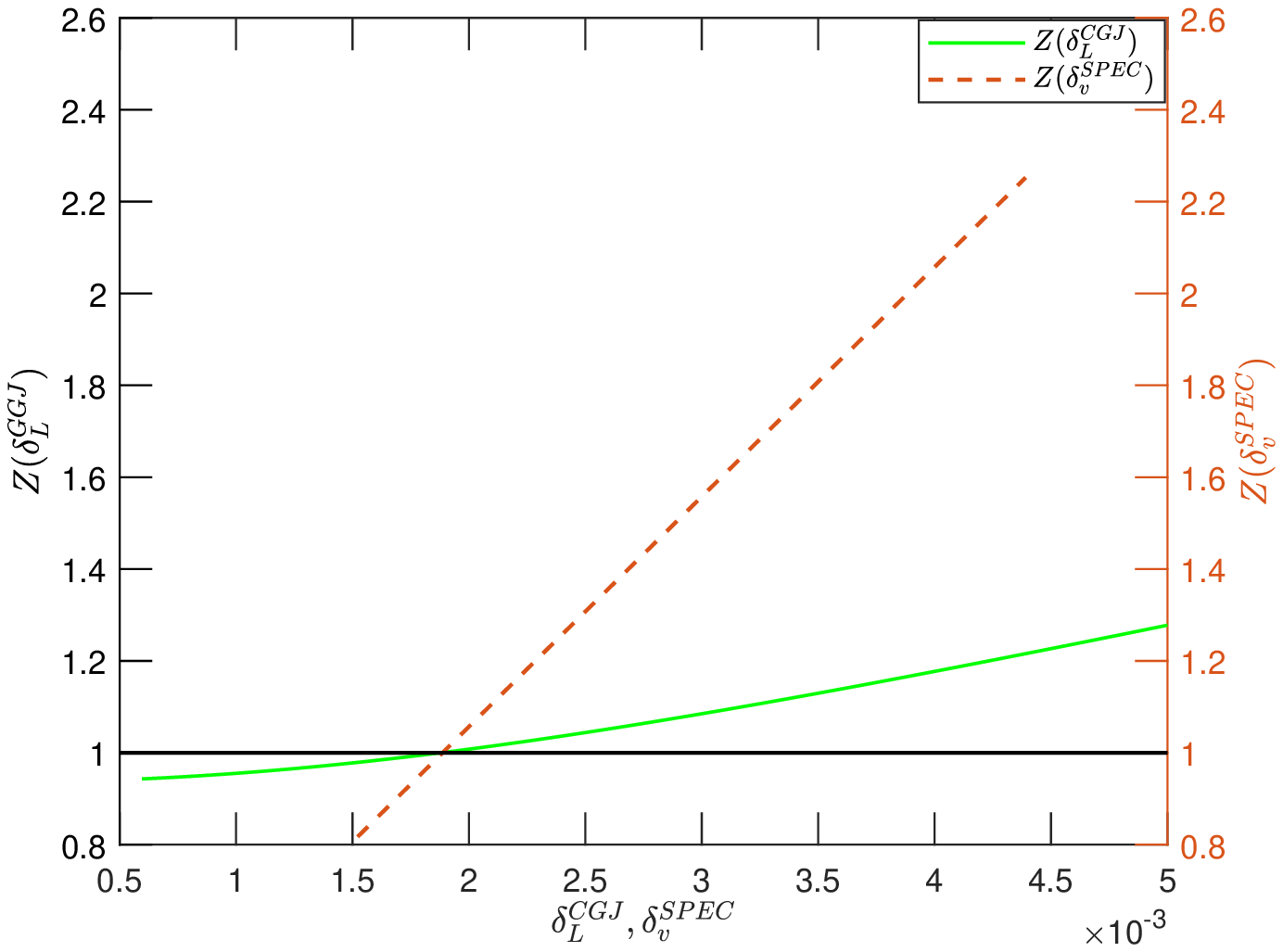}}\par
\subfloat[\label{cyl_tearing_eig}]{\includegraphics[width=9.3cm]{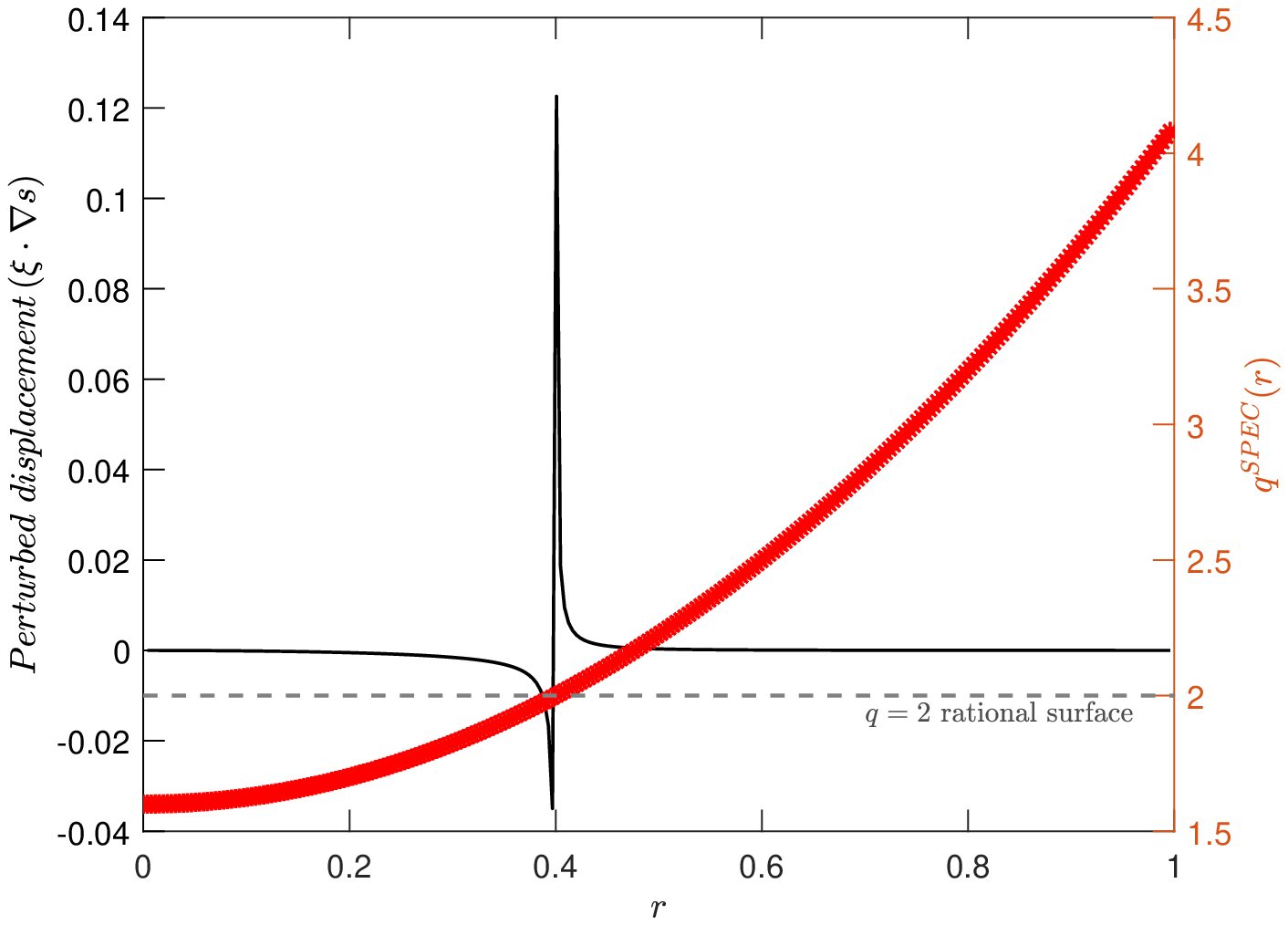}}

\hspace{0mm}
\caption{(a) Solid green line: $Z(\delta_{L}^{CGJ})$ vs $\delta_{L}^{CGJ}$ obtained from Eqn.\eqref{eq.8} computed with the constant plasma density $\rho=1$. We consider the $\mu_{0}=1$ and the Alfven speed $v_{A} = ||\vec{B_{eq}}||/\sqrt{\rho_{0}\mu_{0}}$ in SI units, where $\vec{B_{eq}}$ is the equilibrium magnetic field. It follows that, in our units system $v_{A}$, $\tau_{A}$ and $r$ are unity, and thus $\eta=S^{-1}$, where $S$ is the Lundquist number. Dashed red line: $Z(\delta_{v}^{SPEC})$ vs $\delta_{v}^{SPEC}$
 computed from SPEC. The solid black line indicates the marginal stability threshold condition $Z=1$ ; (b) the SPEC computed perturbed surface displacement ($\boldsymbol{\xi}\cdot\nabla s$) vs $r$ for the $m/n = 2/1$  unstable mode. The SPEC $q$-profile is computed from Eqn.\eqref{eqn.q.cycl}. Here, $N_{v} =180$ is considered.} 
\end{figure}

We investigate the role of resistive singular layer in both models by considering the equilibrium configuration in the scenario of fixed $p_{0}=0.25$ (see Figure \ref{q_cycl}). In Figure \ref{cyl_tearing}, we have plotted $Z(\delta_{L}^{CGJ})$ as a function of $\hat{\delta}$. 
We see that as  $\delta^{CGJ}_{L}$ increases, $Z(\delta^{CGJ}_{L})$ crosses the stability threshold line ($Z=1$) and predicts instability for the mode perturbation $m/n=2/1$. Here, the stability threshold in terms of $\delta_{L}^{CGJ}$ is approximated as $1.77\times 10^{-3}$. Now, the eigenvalues $\lambda_{SPEC}^{2,1}$ from Eqn. \eqref{eq.hess.45} are evaluated numerically using the SPEC code for different values of $\delta_{v}^{SPEC}$, which is understood in terms of Eqn.\eqref{deltaSPEC}. We observed that, for the sufficiently small value of $\delta_{v}^{SPEC}$,  $Z(\delta_{v}^{SPEC})$ crosses its the stability threshold, and coincides 
$Z(\delta_{L}^{CGJ}) > 1$  (see figure \ref{cyl_tearing}). Since the  $m/n=2/1$ mode is unstable, this indicates that the effect of the width of the singular layer in both models is comparable, close to the marginal stability locus. Finally, the Figure \ref{cyl_tearing_eig} shows the radial structure of the SPEC eigenfunction $\boldsymbol{\xi}\cdot\nabla s$ for the $m/n=2/1$ unstable equilibrium case with $\delta_{v}^{SPEC}\sim1.87 \times 10^{-3}$. In the vicinity of the $q=2$ rational surface, a typical spatial behaviour of this tearing eigenfunction can be observed. Here, the $\boldsymbol{\xi}\cdot\nabla s$ is defined as the radial perturbed component of the interface displacement. These results show the marginal stability threshold of CGJ and MRxMHD theory coincide, when the $\delta_{v}^{SPEC}$ is proportional to the $\delta^{CGJ}_{L}$, that is, $\delta_{v}^{SPEC}\sim \delta_{L}^{GGJ}$. 
\par
Moreover, we postulate that as  $\delta_{v}^{SPEC}$ decrease, the ``\textit{ideal interfaces }'' surrounding the $\iotabar=1/2$ resonant surface came sufficiently close to rational surfaces, and induces shielding currents. That is, in the limit of vanishing width of the resonant volume $\delta_{v}^{SPEC}\rightarrow{} 0$, the parallel current density becomes infinite, such that the parallel current within the volume region becomes finite and non-zero. In accordance, the emergence of shielding currents from the `` \textit{ideal interfaces}'' could also be a potential reason for stabilization of this resistive mode. It is the current sheet that allows the small solutions on either side of a singular surface to be disconnected in ideal MHD, screening one side from the other. In ideal MHD, field-line reconnection is forbidden by the frozen-in flux condition, so the current sheet must form to prevent the tearing mode island (of arbitrarily small amplitude in the linearized approximation) that forms in relaxed MHD. This phenomenon has also been observed in the formation of current-sheets in magnetic reconnection \citep{boozer,dbis}. 
\par 
In the vicinity of a resonant volume, SPEC allows a transition from partial Taylor relaxation to ideal MHD as $\delta_{v}^{SPEC}\rightarrow{}0$. Here, it should be noted that the ``\textit{ideal interfaces }'' will not overlap even when SPEC computes an equilibrium solution as $\delta_{v}^{SPEC}$ decreases in the vicinity of the resonant surface. Overlapping of the ``\textit{ideal interfaces }'' are not allowed on both conceptual and computational grounds. 
\par 




\begin{figure}[ht]
\centering

\hspace{0mm}

\end{figure}

\section{Modified tearing mode in large aspect ratio axisymetric plasma  }\label{sec3}

The preceding section showed how the CGJ and MRxMHD model can are related in cylindrical geometry. 
This section compares the Glasser- Greene-Johnson (GGJ) compressible resistive layer model to the variational energy principle of the MRxMHD for the modified tearing modes in a large aspect ratio toroidal geometry. 
\par

The modified tearing instability normally occurs during the course of tokamak discharges due to thermal instability in Ohmic plasmas \citep{frs1970}. Usually, the stabilising effect of magnetic shear on ideal interchange instability is eliminated by these modified tearing modes. Glasser-Greene-Johnson were the first to investigate the tearing stability threshold in relation to shear, pressure, and toroidicity.

 Our primary goal in this section is to investigate the presence of the resistive volume layer width within MRxMHD, in comparison to the classic GGJ model. In GGJ, the resistive width of a singular layer (here, we denoted as $\delta^{GGJ}_{L}$) depends on the plasma inertia (growth rate of an unstable
ideal MHD mode) and the plasma resistivity. It approaches zero as the growth rate or the resistivity reduces to zero simultaneously. 
\subsection{Stability condition between GGJ and MRxMHD model}\label{sec3a}

In GGJ model \cite{GGJ1976}, while an equilibrium configuration is ideally stable for $q_{0}>1$ and satisfies the resistive interchange stability condition $D_{R}<0$, it is unstable to a special case of tearing mode known as modified tearing mode if $\Delta' > \Delta_{crit}$ in the vicinity of resonant location. Here, we define  \begin{eqnarray} 
D_{R} &=& \frac{2q^{4} p'}{rB_{0}^{2} q'^{2}} \left(\!\!1-\frac{1}{q^2} +\frac{qq'}{r^3}\right) \!\!\int_{0}^{r} \!\!dt\!\! \left(\frac{t^{3}}{q^2}\!\! - \!\!\frac{2 R_{0}t^{2}p'}{B_{0}^{2}}\right)\!, \\
    \Delta_{crit}&\approx& 1.54 \left(\frac{V_{S}}{X_{0}}\right)|D_{R}|^{5/6},
\end{eqnarray} and $\Delta'$ being the generalized jump in the logarithmic derivative of the perturbed magnetic flux across the resistive layer. Here, $V_{s}/X_{0}$ is defined as the ratio of the macrosopic resistive scale length to a magnetohydrodynamic scale that varies as $\eta^{1/3}$, \par

 To examine the stability of a modified tearing mode in the vicinity of singular layer where the magnetic winding number is rational, we introduce a localized stability parameter $Z(\delta^{GGJ}_{L},\upsilon) $ which is a function of both $\delta^{GGJ}_{L}$ and a magnetic shear parameter  $\upsilon \in \mathbf{Z}^{+}$. We define the $Z(\delta^{GGJ}_{L},\upsilon)$ as, \begin{eqnarray}
    Z(\delta^{GGJ}_{L},\upsilon) = \frac{\Delta'}{\Delta_{crit}}  , 
\end{eqnarray} where $\Delta_{crit}$ is re-written in terms of the resistive width of a singular layer $\delta^{GGJ}_{L}$ as,\begin{equation}
    \Delta_{crit}\approx \frac{1.54\,|D_{R}|^{5/6}|Q|^{1/4}}{\delta^{GGJ}_{L}},
\end{equation} with $Q$ being a complex and dimensionless variable. It is obtained by solving a dispersion relation when the boundary layer solutions of inner layer are matched with the outer layer. This dispersion relation is written as \begin{equation}\label{eq.7}
    \Delta(Q)=\Delta',
\end{equation} where
\begin{eqnarray}\label{eqn24}
    &\Delta'(Q)=&\frac{\pi D_{R}(2V_{s}/X_{0})^{1-2H} \Gamma(1/4)\Gamma^{2}(1-{H}/{4})}{(1-2H)\Gamma(1-{H}/{2})} \\
    &\times& \frac{\Gamma(3/4-H/2)Q^{{(2H-1)}/{4}}}{\left(\cos({\pi H}/{2})\Gamma({1+H}/{2})\Gamma(1-H)\right)^{2}}\nonumber\\
    &\times& \left( \frac{Q^{3/2}}{D_{R}}-\frac{\,\Gamma(3/4)\Gamma^{2}(1/2-H/4)\Gamma(1/4-H/2)}{4\Gamma(1/4)\Gamma^{2}(1-H/4)\Gamma(3/4-H/2)}\right).\nonumber
\end{eqnarray} where $\Gamma$ is the gamma function and the analytical expressions for $H$, $V_{s}/X_{0}$ and $\delta^{GGJ}_{L}(r)$ are written as
\begin{eqnarray}
    H &=& - \frac{2q^{5} p'}{r^{4}B_{0}^{2} q'^{}}  \!\!\int_{0}^{r} \!\!dt\!\! \left(\frac{t^{3}}{q^2}\!\! - \!\!\frac{2 R_{0}t^{2}p'}{B_{0}^{2}}\right), \label{H}\\
 \label{vsxo}
    \frac{V_{S}}{X_{0}}&=&\left(\frac{n B_{0}}{\eta R}\frac{q'(r)}{q(r)} \right)^{1/3} (1+2q(r)^{2})^{-1/6}\rho^{-1/6},\\
\label{delta_GGJ}
    \delta^{GGJ}_{L} &=& \left(\left(\frac{\eta R}{n B_{0 }}\frac{q(r)}{q'(r)}\right)^{2}\!\! (1+2q(r)^{2})\right)^{1/6}\!\!|Q|^{1/4}\rho^{1/6}\!\!\!.
\end{eqnarray} 
The above Eqns. \eqref{H}, \eqref{vsxo} and \eqref{delta_GGJ} are available from Glasser \etal\citep{GGJ1976}(see Eqns. (A21), (A31) and (A32)).
\par 
An instability occurs (either pressure-induced or modified tearing mode), if there exists a solution of Eqn.\eqref{eq.7} with $Re(Q) > 0 $, for a given value of $\Delta'$.
The modified tearing mode is unstable if $Z(\delta^{GGJ}_{L},\upsilon)>1$ condition is satisfied, otherwise stable below the critical condition of $Z(\delta^{GGJ}_{L},\upsilon) =1$, which denotes the marginal stability condition for this kind of tearing mode.


 
\par
From the perspective of the MRxMHD model, we now proceed to characterise the axisymmetric equilibria and its stability. In accordance with the previous Sec.\ref{sec2a}, we first let the resonant volume have an arbitrary radial width, denoted by $\delta_{v}^{SPEC}$. Then, the input parameters such as $K_{l}$, $\Delta \psi_{t,l}$, $\Delta \psi_{p,l}$, are determined by discretizing the volume-averaged magnetic helicity $\langle K_{} \rangle$ and the flux-averaged toroidal and poloidal fluxes ($\tilde{\psi}_{tw}$ and $\tilde{\psi}_{pw}$) profiles over the required number of volumes $N_{v}$. These averaged quantities are evaluated from the given equilibrium profiles which are considered for the simulations. Thus, the size of the $\delta_{v}^{SPEC}$ is also parameterized by the $K_{l}$ and the enclosed fluxes. Here, $\langle K_{}\rangle$ is also normalised to the total helicity and the $\tilde{\psi}_{w}$ varies between $0$ and $1$. Then, adjacent to a resonant volume where we determine the KAM surfaces, SPEC's toroidal co-ordinate is constructed as, $ {\bf x}_{l}(\theta,\zeta)=R_{l}(\theta,\zeta)\hat{\bf e}_{R}+Z_{l}(\theta,\zeta)\hat{ \bf e}_{Z}$.
 Here, $\hat{\bf e}_{R}=\cos\phi \hat{\bf i} +\sin\phi \hat{\bf j}$ for the toroidal angle $\zeta\sim\phi$, and the $R_{l}(\theta,\zeta)$ and $Z_l(\theta,\zeta)$ are an even and odd function of $(\theta,\zeta)$ respectively. The symmetric and non-symmetric variables are discretized in the Fourier basis function as $R_l(\theta,\zeta)=\sum_{i} R_{l,i} \cos(m_{i}\theta-n_{i} N_{p}\zeta)$ and $ Z_l(\theta,\zeta)=\sum_{i} Z_{l,i} \sin(m_{i}\theta-n_{i} N_{p}\zeta)$.

\par
In toroidal geometry, the stability of an MRxMHD equilibrium can be assessed by considering the infinitesimal variation of the interface force balance term, $f_{l}=-[[p+B^{2}/2\mu_{0}]]_{l}{\bf n}_{l}$, with respect to the poloidal and toroidal perturbation of interface geometry ${\bf x}_{l}$ \cite{Kumar_2022}. Similar to the previous cylindrical stability implementation in SPEC, this form of change is also numerically interpreted as the Hessian matrix, which can be written as\citep{Kumar_2022} 
\begin{eqnarray}\label{eq.hess.46}
    {\mathbf H}_{j,k,l,{l'}}&=&\frac{\delta}{\delta {\bf x}_{l',k}} \left(\delta F/\delta {\bf x}_{l,j}\right),
\end{eqnarray} where $j$ and $k$ are defined as dummy variable for the Fourier harmonics for clarity, with $N_{m,n}$ being the total number of Fourier modes, and $l$ and $l'$ represent the different interface labels. To include the effects of finite compressibility in toroidal geometry, the pressure variation can be computed using the Eqn.\eqref{pre_Var} where $\delta V_{l}=(\partial V_{l}/\partial R_{l})\delta R_{l}+(\partial V_{l}/\partial Z_{l})\delta Z_{l}$.
 The expression for $V_l$ can be obtained by the integral
\begin{eqnarray}
 V_{l} &=& \int_{{ \Omega_l}} d^{3}\tau = \frac{1}{3}\int_{{ \Omega_l}} \; \nabla \cdot {\bf x}_{l}\,\, d^{3}\tau = \frac{1}{3} \int_{{ \delta \Omega_l}} \; {\bf x}_{l} \cdot d{\bf S}, \\
 &=& \frac{1}{3} \int_{0}^{2\pi} d{\theta} \int_{0}^{2\pi/N} d{\zeta} ({\bf x}_{l} \cdot {{\bf e}_{\theta}} \times 
{{\bf e}_{\zeta}}),\\
   &=&  \frac{1}{3} \int_{0}^{2\pi}\!\!\!d{\theta} \int_{0}^{2\pi/N} d{\zeta} \; R_{l} \left( Z_{l} R_{l,\theta} - R_{l} Z_{l,\theta}  \right) \label{eqn.30},
 \end{eqnarray} where we have considered $\nabla \cdot {\bf x}_l = 3$. On expanding the Eqn.\eqref{eqn.30} as a summation of the Fourier harmonics, we have 
\begin{eqnarray}
\begin{aligned}
    V_{l}  & =  \frac{1}{3}  \; \sum_i \sum_j \sum_k R_{l,i} \left(Z_{l,j} R_{l,k} - R_{l,j} Z_{l,k} \right) (+m_k)\\ \times& 
             \oint \!\!\!\! \oint \!\! \; \cos\alpha_i \cos\alpha_j \cos\alpha_k \,d\theta d\zeta ,
\end{aligned}
\end{eqnarray}
where $i^{th}$,$j^{th}$ and $k^{th}$ are the Fourier harmonics of $R_{l},Z_{l}$. Then, the partial derivatives $\frac{\partial V_{l}}{\partial R_{l,i}}$ and $\frac{\partial V_{l}}{\partial Z_{l,i}}$ are obtained as
\begin{align}
   3  \frac{\partial V_{l}}{\partial R_{l,i}}  = & 
       \left(Z_{l,j} R_{l,k} m_k - R_{l,j} Z_{l,k} m_k - R_{l,j} Z_{l,k} m_k \right) \\ 
       &\times \oint\!\!\!\!\oint \!\! d\theta d\zeta \; \cos\alpha_i \cos\alpha_j \cos\alpha_k \nonumber\\
        + & \left( - Z_{l,j} R_{l,k} m_k + R_{l,j} Z_{l,k} m_k + R_{l,j} Z_{l,k} m_k \right)\nonumber \\ 
    & \times \oint \!\!\!\! \oint \!\! d\theta d\zeta \; \cos\alpha_i \sin\alpha_j \sin\alpha_k, \nonumber
      \end{align}
     
and 
 \begin{align}
 3 \frac{\partial V_{l}}{\partial Z_{l,i}} = & \left( - R_{l,k} R_{l,j} m_i                                             \right)  
                                                            \oint \!\!\!\! \oint \!\! d\theta d\zeta \; \cos\alpha_i \cos\alpha_j \cos\alpha_k
                                                     \\  +&  \left( - R_{l,k} R_{l,j} m_{k}                                             \right) 
                                                            \oint \!\!\!\! \oint \!\! d\theta d\zeta \; \cos\alpha_i \sin\alpha_j \sin\alpha_k.\nonumber
    \end{align}

 When this matrix $\mathbf{H} $ is evaluated at fixed magnetic helicity and enclosed fluxes, its eigenvalues provide information about the stability corresponds to each Fourier mode harmonics $m,n$. These eigenvalues, $\lambda^{m,n}_{SPEC}$ from $\mathbf{H}$ are evaluated numerically using the SPEC-Hessian calculation. 
 
 \par
Following Eqn.\eqref{deltaSPEC} of Sec.\ref{sec2b}, we now express the smallest negative eigenvalue (normalized to its maximum value), referred as $min\{\lambda^{m,n}_{SPEC}\}$ in terms of the stability parameter $Z(\delta_{v}^{SPEC},\upsilon)$ as,\begin{eqnarray}\label{eq35}
    Z(\delta_{v}^{SPEC},\upsilon)= 1- min\{\lambda^{m,n}_{SPEC}(\delta_{v}^{SPEC},\upsilon)\},
\end{eqnarray}
 such that the stability conditions of $Z(\delta^{GGJ}_{L},\upsilon)$ can be interpreted same as described before.

\subsection{Equilibrium and stability conditions}\label{sec3b}
 In this section, we discuss the modified tearing instability of a model circular tokamak (large aspect ratio ) equilibrium considered in Glasser \etal \citep{GGJ1976}. The aspect ratio $A=8.4$, where the major radius $R_{0}=8.4 \,m$ and minor radius $a=1\, m$, such that all the equilibrium scalars are independent of the toroidal angle $\phi$ about the axis of symmetry. The equilibrium toroidal current density and the parabolic pressure profiles are described as a function of $r$, which is
\begin{eqnarray}
J_{\phi}(r)=J_{0}/(1 + \upsilon \,r^{2}/a^{2})^{2},\hspace{4mm}
p(r)=p_{0}(1-r^{2}/a^{2}),
\end{eqnarray} where $\upsilon \in \mathbf{Z}^{+}$ is the shear parameter, $J_{0}=2B_{0}/q_{0}R_{0}$ and $p_{0}=\beta_{p}(B_{0}a/q_{0}R_{0}(1+\upsilon))^{2}$. The analytical expression for poloidal plasma beta is obtained as $
    \beta_{p}=\left(\frac{2Rq(a)}{B_{0}a^{2}}\right)^{2}\int_{0}^{a} rp(r) \,dr$. We investigate this equilibrium model in the scenario with $q_{0}=1.1$, $\beta_{p}=0.8$ and shear parameter values $\upsilon=2$ and $\upsilon=3$. This shear parameter $\upsilon$ plays a critical role in destabilizing and stabilizing factor of this plasma configuration.

 \begin{figure}[ht]
\centering
\subfloat[\label{fig21}]{
  \includegraphics[width=9.0cm]{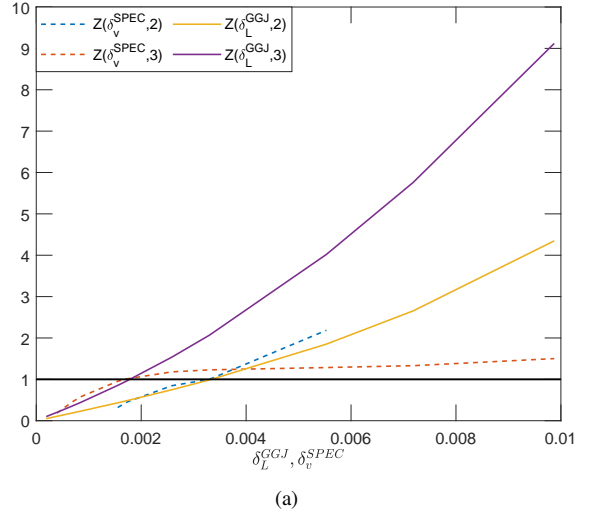}
}\par
\subfloat[\label{toktear_ig}]{
  \includegraphics[width=9.3cm]{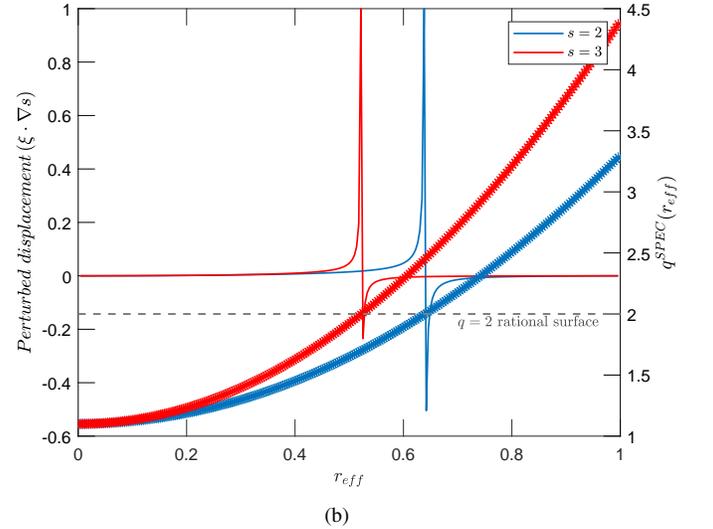}
}

\hspace{0mm}
\caption{(a) $Z(\delta_{v}^{SPEC},\upsilon)$ and $Z(\delta^{GGJ}_{L},\upsilon)$ as a function of $\delta_{v}^{SPEC}$ and $\delta^{GGJ}_{L}$, respectively for shear parameter values $\upsilon=2$ and $\upsilon=3$. To obtain $Z(\delta^{GGJ}_{L},\upsilon)$, we consider the mass density $\rho= 1$, $\mu_{0}=1$ and the Alfven speed $v_{A} = ||\vec{B_{eq}}||/\sqrt{\rho_{0}\mu_{0}}$ in SI units, where $\vec{B_{eq}}$ is the equilibrium magnetic field. It follows that, in our units system $v_{A}$, $\tau_{A}$ and $r$ are unity, and thus $\eta=S^{-1}$ where $S$ is the Lundquist number. The solid black line indicates the marginal stability threshold condition $Z=1$; (b) On left axis: the SPEC computed perturbed surface displacement ($\boldsymbol{\xi}\cdot\nabla s$) vs $r_{eff}\sim\psi_{t}/\psi_{edge}$ for unstable $m/n = 2/1$ mode; On right axis : the SPEC computed $q$ profile vs  $r_{eff}\sim\psi_{t}/\psi_{edge}$.     } 
\end{figure}
 
    Figure \ref{fig21} compares the results of $Z(\delta_{v}^{SPEC},\upsilon)$ and $Z(\delta^{GGJ}_{L},\upsilon)$ as a function of  $\delta_{v}^{SPEC}$ and $\delta^{GGJ}_{L}$, respectively. The eigenvalues $\lambda_{SPEC}^{2,1}$ from Eqn. \eqref{eq.hess.46} are evaluated numerically using the SPEC
code for different values of the $\delta_{v}^{SPEC}$, which is understood in terms of Eqn.\eqref{eq35}. For $\upsilon=2$, we observe that the $Z(\delta_{v}^{SPEC},2)$ predicts instability as it crosses the marginal stability threshold line ($Z=1$), for sufficiently small value of $\delta_{v}^{SPEC}$, which is approximated as 2.8 $\times$10$^{-3}$. In addition, the value of $\delta_{L}^{GGJ}$ at which the $Z(\delta^{GGJ}_{L},2)$ crosses its stability threshold, coincides with that at which $Z(\delta_{v}^{SPEC},2)$ is greater than 1. This confirms the potential relationship $\delta_{v}^{SPEC} \sim \delta_{L}^{GGJ}$. Now for $\upsilon=3$, it is observed that $Z(\delta_{v}^{SPEC},3)$ predicts instability for the smaller value of $\delta_{v}^{SPEC}$ than the case of $\upsilon=2$. This is because as $\upsilon$ increases, the current channel shrinks in the vicinity of the rational surface, and the both $\delta_{v}^{SPEC}$ $\delta_{L}^{GGJ}$ reduces. For $\upsilon=3$, the threshold $\delta_{v}^{SPEC}$ is approximated as 1.72$\times$10$^{-3}$, and a similar threshold behavior is found for $Z(\delta_{L}^{GGJ},3)$. Thus, the MRxMHD stability boundary is in agreement with the linear modified tearing mode theory. 
However, if $\delta_{v}^{SPEC}$ becomes sufficiently large compared to $\delta^{GGJ}_{L} $, it can be conceptualized that the pressure flattening in SPEC can indeed remove the stabilizing effects and
considerably affect the stability boundary of the mode. For $p_{l}=const.$ over a larger volume width, the mode can still be strongly destabilized in SPEC and finds different stability threshold. 

Finally, Figure \ref{toktear_ig} shows the
spatial structure of the SPEC eigenfunction $\boldsymbol{\xi}\cdot \nabla s$ as a function of effective radius $r_{eff.}$. The corresponding $m/n=2/1$ unstable equilibrium for $\nu=2$ and $3$ are considered with $\delta_{v}^{SPEC}$ equal to $3.5\times 10^{-3}$ and $2.5 \times 10^{-3}$, respectively. We would like to remark that the SPEC-stability results shown in Fig. 2 are converged in the sense that increasing the Fourier resolution and the radial basis function. Here $N_{v}= 180$ is considered.
\section{Conclusion and future work}\label{sec4}

In this article, we have investigated the impact of the variational energy principle of the MRxMHD model to predict the finite-pressure linear tearing stability of tearing modes. For low pressure plasma, we have investigated a technique with which we have been able to establish a relationship between
the resistive singular layer theories of CGJ, GGJ, and the MRxMHD model. Our analyses shows that the SPEC shows the stabilizing effects as the width of resistive volume layer is decreased. Indeed, if $\delta_{v}^{SPEC}\sim \delta^{CGJ}_{L}$ and  $\sim \delta^{GGJ}_{L}$ {that is,} the effects of finite resistivity and pressure-gradient roughly compensate, and the overall marginal stability of the mode is same for SPEC, CGJ and GGJ. Physical insights into the spatial structure of eigenfunction of the pressure-driven tearing modes computed from SPEC, have clarified the applicability regime of MRxMHD model. Our results indicates the possibility to couple the MATCH code \citep{2014APS..DPPBP8044G}, which solves the resistive inner layer equations in toroidal geometry, with SPEC, not only to predict the stability of MRxMHD plasma, but also to approximate the growth rates (quantitatively) .
\par
In addition to these studies, we anticipate that it may also be possible to establish a relationship between the pressure flattening model (in the vicinity of resonant rational surfaces) discussed in Refs. \citep{Ham_2012,Bishop_1991} with our model. As a matter of fact, as pressure increases, it is commonly observed that the Mercier indices move apart and it becomes difficult to obtain the large and small solutions in the vicinity of the rational surface \citep{Fitzpatrick_1993,Ham_2012}. 
This restriction can be overcome in both MRxMHD and Ham \etal \citep{Ham_2012} model, due to the pressure flattening at the rational surface. We therefore aim to address this in our future investigations. 

When using MRxMHD to predict nonlinear tearing mode saturation; the difference between the potential energy corresponding to the equilibrium and the secondary minimized total energy can be interpreted as the second variation in the nonlinear stability case. 
Following this, in slab geometry, Loizu \etal \citep{loizunon} demonstrated that the nonlinear saturation of tearing modes can be predicted directly with SPEC using appropriate constraints, without resolving the complex resistivity-dependent dynamics and without free parameters. To extend Loizu's work for finite beta cylindrical or toroidal plasma, our technique to compute an initial unstable MRxMHD equilibrium state can be utilized. We intend to investigate this work further in the future.
\par


\par
\section*{Acknowledgement}
The first author (A.K) would like to acknowledge the Stellaratortheorie Dept., Max-Planck-Institut fuer Plasmaphysik, for its hospitality during the writing of the latter part of the article. 
We acknowledge the support by the Australian Research Council project No.DP170102606, Simons Foundation grant SFARI No.560651/A.B. J.L works carried out within the framework of the EUROfusion consortium and has received funding from the Euratom research and training programme 2014-2018 and 2019-2020 under GA No.633053. The views and opinions expressed herein do not necessarily reflect those of the European Commission.
This computational research is undertaken with the assistance of resources and services from the National Computational Infrastructure (NCI), which is supported by the Australian Government in the framework of ANU Merit Allocation Scheme.

\nocite{*}
\bibliography{aipsamp}

\end{document}